\documentclass[conference]{IEEEtran}
\usepackage{cite}
\usepackage{amsmath,amssymb,amsfonts}
\usepackage{algorithmic}
\usepackage{graphicx}
\usepackage{caption}
\usepackage{subcaption}
\usepackage{textcomp}
\usepackage{booktabs}
\usepackage{float}
\usepackage{url}
\usepackage{makecell}
\usepackage{xcolor}
\def\BibTeX{{\rm B\kern-.05em{\sc i\kern-.025em b}\kern-.08em
    T\kern-.1667em\lower.7ex\hbox{E}\kern-.125emX}}

\makeatletter
\newcommand{\linebreakand}{%
  \end{@IEEEauthorhalign}
  \hfill\mbox{}\par
  \mbox{}\hfill\begin{@IEEEauthorhalign}
}
\makeatother

\begin{document}

\title{Towards an Adaptive Runtime System for Cloud-Native HPC}

\author{
  \IEEEauthorblockN{Aditya Bhosale}
  \IEEEauthorblockA{\textit{University of Illinois Urbana-Champaign}\\
  Urbana, IL \\
  adityapb@illinois.edu}
  \and
  \IEEEauthorblockN{Advait Tahilyani}
  \IEEEauthorblockA{\textit{University of Illinois Urbana-Champaign}\\
  Urbana, IL \\
  advaitt3@illinois.edu}
  
  \linebreakand 

  \IEEEauthorblockN{Laxmikant Kale}
  \IEEEauthorblockA{\textit{University of Illinois Urbana-Champaign}\\
  Urbana, IL \\
  kale@illinois.edu}
  \and
  \IEEEauthorblockN{Sara Kokkila-Schumacher}
  \IEEEauthorblockA{\textit{IBM Thomas J. Watson Research Center}\\
  Yorktown Heights, NY \\
  saraks@ibm.com}
}

\maketitle

\begin{abstract}
The ongoing convergence of HPC and cloud computing presents a fundamental challenge: HPC applications, designed for static and homogeneous supercomputers, are ill-suited for the dynamic, heterogeneous, and volatile nature of the cloud. Traditional parallel programming models like MPI struggle to leverage key cloud advantages, such as resource elasticity and low-cost spot instances, while also failing to address challenges like performance variability and processor heterogeneity.
This paper demonstrates how the asynchronous, message-driven paradigm of the Charm++ parallel runtime system can bridge this gap. 
We present a set of tools and strategies that enable HPC applications to run efficiently and resiliently on dynamic cloud infrastructure across both CPU and GPU resources. 
Our work makes two key contributions. First, we demonstrate that rate-aware load balancing in Charm++ improves performance for applications running on heterogeneous CPU and GPU instances on the cloud. We further demonstrate how core Charm++ principles mitigate performance degradation from common cloud challenges like network contention and processor performance variability, which are exacerbated by the tightly coupled, globally synchronized nature of many science and engineering applications. Second, we extend an existing resource management framework to support GPU and CPU spot instances with minimal interruption overhead. Together, these contributions provide a robust framework for adapting HPC applications to achieve efficient, resilient, and cost-effective performance on the cloud.

\end{abstract}


\section{Introduction}

The historical divergence between HPC and cloud computing is rapidly closing. Traditionally, HPC systems have prioritized raw performance for tightly-coupled scientific simulations on static, homogeneous hardware. On the other hand, cloud platforms were built for resilience, elasticity, and resource dynamism to serve loosely-coupled workloads like web services.

Today, this gap is closing due to several factors. The accelerating pace of hardware innovation, the massive capital investment required for next-generation systems, and the exponential growth of AI/ML workloads are driving HPC users towards the cloud. Similarly, cloud providers are now offering specialized hardware, such as dedicated HPC instances and high-performance networks, to attract these HPC users. This convergence presents a significant opportunity to democratize access to high-end computing but also exposes fundamental challenges at the intersection of these two paradigms.

In parallel with this trend, the HPC landscape has fundamentally shifted from traditional CPU-only architectures to heterogeneous systems where GPUs are indispensable. Modern scientific and machine learning workloads, from molecular dynamics to deep learning, now rely on the massive parallel processing capabilities of GPUs to achieve performance unattainable with CPUs alone.

This convergence of HPC and the cloud raises the question of how applications designed for the static, predictable world of traditional HPC can run efficiently in the dynamic and sometimes unreliable cloud environment. Key challenges in adapting HPC to the cloud include network performance variability, hardware heterogeneity, and fault tolerance. Traditional parallel programming models like MPI are not well-suited to address these challenges. Alternative programming models like Charm++ have emerged as an efficient model for execution on cloud platforms~\cite{Gupta:Cloudcom2013, malleable2014}. Previous work from Bhosale et al.~\cite{Bhosale2025Cloud} demonstrated the use of Charm++ for running HPC applications on preemptible discounted spot instances.

While the previous work shows significant cost savings from using spot instances, it revealed three major limitations. 
First, the previous work failed to address issues of network contention, and hardware heterogeneity and performance variability commonly found in cloud platforms. 
Second, the previous work was limited to CPUs. Given the changing landscape of HPC, to be a truly effective runtime system for cloud-native HPC, it needs to manage both CPU and GPU workloads efficiently.
Third, the temporary drop in capacity while handling spot interruptions had a significant effect on the overall runtime of the application, which can be a problem especially for deadline-sensitive applications.

In this paper, we address these three limitations through the following contributions. First, we demonstrate that the core principle of overdecomposition in Charm++ efficiently mitigates performance degradation caused by network contention (Section~\ref{sec:network}).
We show that rate-aware dynamic load balancing enables applications to effectively utilize heterogeneous cloud instances (Section~\ref{sec:hetero}). By applying this strategy to two applications without inherent load imbalance, we achieved up to 25\% performance improvement when running on a mix of different instance types.

Second, we extend the \texttt{CharmCloudManager} framework~\cite{Bhosale2025Cloud} to support GPU workloads by managing interruptions on GPU spot instances with minimal overhead (Section~\ref{sec:spot:gpu}).

Third, we extend \texttt{CharmCloudManager} to proactively launch replacement instances when spot nodes are at an elevated risk of interruption. This approach minimizes capacity drops and reduces the overhead of handling spot preemptions (Section~\ref{sec:spot:rebalance}). Our results show that a proactive replacement policy reduces the overhead of handling interruptions by 50\% compared to a reactive approach used in the previous work~\cite{Bhosale2025Cloud}. We further show that the end-to-end application runtime overhead caused by interruptions is reduced from 16\% and 20\% down to less than 1\% and 3.4\% for CPU and GPU instances, respectively.

\section{Background}

\subsection{Charm++}

Charm++~\cite{sc14charm, Charm++} is an asynchronous, message-driven parallel programming model where computation is expressed in terms of C++ objects or chares. These chares are mapped to processing elements (PEs) by the Charm++ runtime system. The chares can communicate with each other via non-blocking entry method invocations which are delivered by the Charm++ runtime using a distributed location management system.

Chares can be organized into one or more indexed collections. An individual chare in a collection is identified (or named) by its collection name and an index within it. All chares in a given collection are instances of the same class, and the system supports broadcasts and reductions over each collection.  

This object-based design has two key principles:

\subsubsection{Overdecomposition}

Charm++ applications are broken down into a number of chares larger than the number of available processors. When a processor is waiting on a remote message, the runtime can schedule other available chares on that processor, effectively hiding the communication latency by overlapping it with useful computation.

\subsubsection{Migratable Objects}

The fine-grained decomposition of Charm++ applications allows the runtime to treat chares as migratable units of work. 
Since the runtime system manages the location of each chare, it can dynamically migrate them between PEs during execution. This migratability enables features such as dynamic load balancing, where the runtime can move objects from overloaded to underloaded PEs to improve application performance based on object and PE load data collected by the runtime system.

\subsection{Resource Elasticity in Charm++}

The migratability of objects also enables the runtime system to move objects to change the number of PEs of an application at runtime~\cite{malleable2014}. To scale down the number of PEs (\textit{shrink}), the runtime system moves objects away from the processors to be removed. The application state is then checkpointed in Linux shared memory, and the application is restarted with a smaller number of PEs. Similarly, to scale up the number of PEs (\textit{expand}), the application state is first checkpointed in Linux shared memory. The application is then restarted with the new number of PEs, and a load balancing step is conducted to balance the load across all PEs.

\section{Performance Challenges of HPC on Cloud}

\subsection{Poor Network Performance}
\label{sec:network}

Standard cloud networking relies on the TCP/IP protocol, which requires the operating system's kernel to manage connections, package data, and handle reliability. This process adds significant latency to every message. 
Furthermore, multi-tenancy in cloud instances can cause high latency due to network contention.
Recognizing this gap, several cloud providers have recently built dedicated solutions for HPC applications. On Amazon Web Services\textsuperscript{\textregistered} (AWS), for example, the key technology is the Elastic Fabric Adapter (EFA)~\cite{aws-efa}.

However, EFA is typically available only on larger instance sizes, which have lower spot availability~\cite{spotlake}. Therefore, to use the cloud effectively for HPC, a cloud-native runtime system must be able to operate efficiently even on the standard, high-latency networks of more readily available instances.

\begin{figure}
    \centering
    \includegraphics[width=\linewidth]{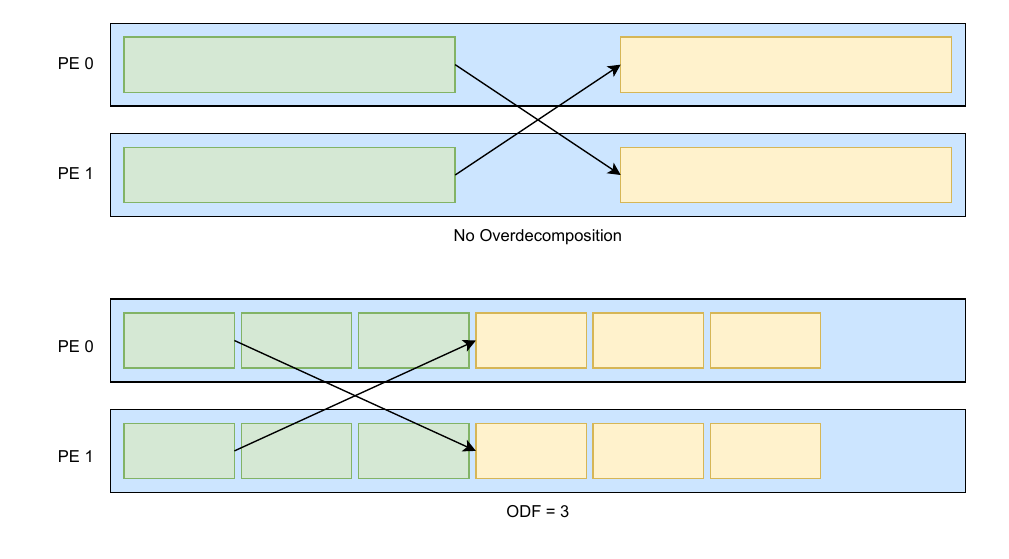}
    \caption{Communication latency hiding using overdecomposition in Charm++}
    \label{fig:overdecomposition}
\end{figure}

Charm++ uses overdecomposition as one of its core principles. Overdecomposition can hide communication latencies by overlapping communication with computation from other chares as shown in Figure~\ref{fig:overdecomposition}.

\subsubsection{Experiments}

To evaluate the effectiveness of overdecomposition at hiding communication latency, we use the Jacobi2D example. This application solves the Laplace equation on a 2D grid by iteratively calculating the average of its four neighbors at each point on the grid. Although simple, this benchmark captures some essential features of HPC/CSE workloads, including neighbors-only communication, less frequent global synchronizations, and communication-computation tradeoffs. 
We ran a weak scaling experiment on CPUs and GPUs for the example with different overdecomposition factors and measured the average time per iteration. The specific instance types and hardware configurations used for these experiments are summarized in Table~\ref{tab:instance-details}.

\begin{figure*}
    \centering
    \begin{subfigure}[b]{.45\linewidth}
        \centering
        \includegraphics[width=\linewidth]{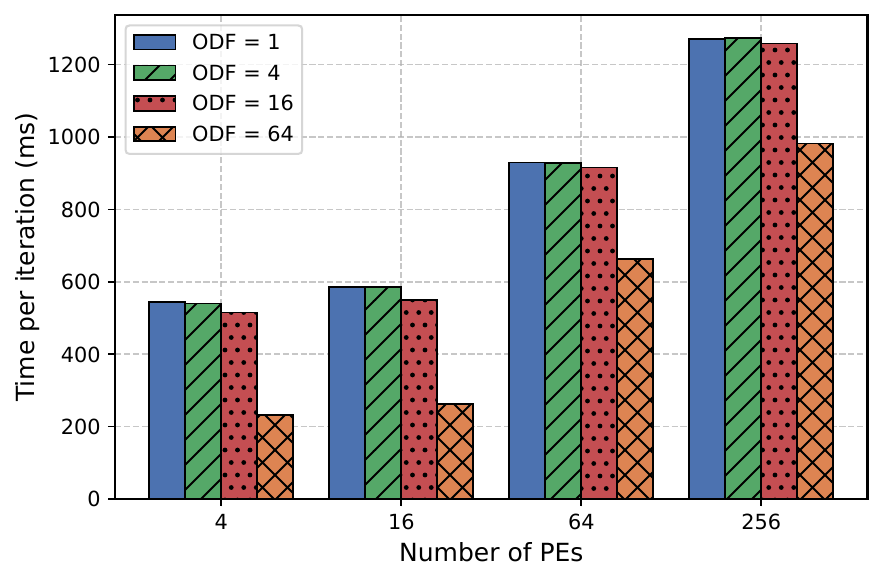}
        \caption{Weak scaling on CPUs}
        \label{fig:weak-scaling:cpu}
    \end{subfigure}
    \begin{subfigure}[b]{.45\linewidth}
        \centering
        \includegraphics[width=\linewidth]{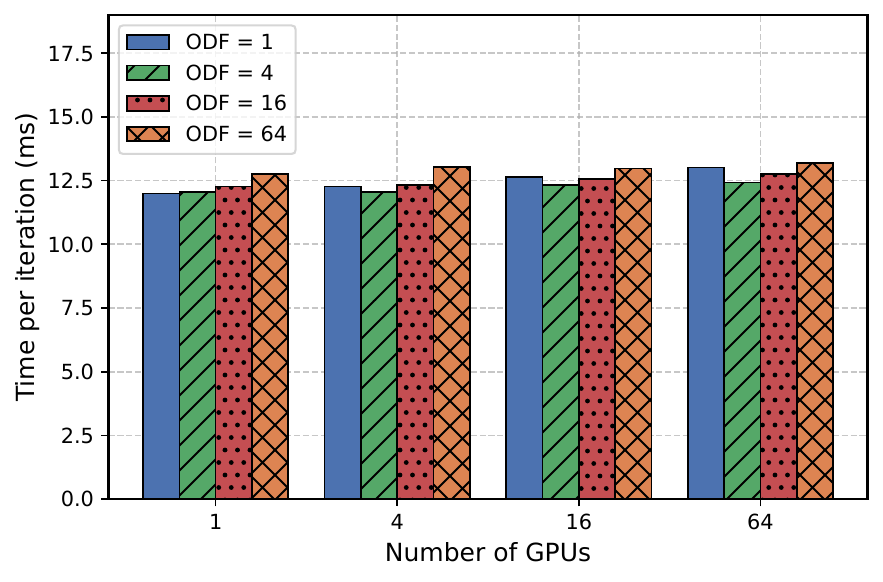}
        \caption{Weak scaling on GPUs}
        \label{fig:weak-scaling:gpu}
    \end{subfigure}
    \caption{Weak scaling for Jacobi2D with different overdecomposition factors}
    \label{fig:weak-scaling}
\end{figure*}

\begin{table}[h]
\centering
\caption{Instance configurations for CPU and GPU experiments on AWS.}
\label{tab:instance-details}
\begin{tabular}{@{}lll@{}}
\toprule
\textbf{Category} & \textbf{Instance Type} & \textbf{Hardware Details (per Instance)} \\ \midrule
\textbf{CPU} & c6a.2xlarge & 4 Physical Cores (AMD EPYC) \\
             & c6a.4xlarge & 8 Physical Cores (AMD EPYC) \\ \midrule
\textbf{GPU} & g4dn.xlarge & 4 vCPUs, 1 NVIDIA T4 GPU (16 GB) \\ \bottomrule
\end{tabular}
\end{table}

For the experiment on CPUs, we used c6a.2xlarge instances with 4 physical cores each for up to 64 PEs, and c6a.4xlarge with 8 physical cores each for larger number of PEs. We used a grid size of $8192 \times 8192$ per PE. We used the \texttt{ucx-linux-x86} build of Charm++. We manually tiled the computation loop to the same tile size for all overdecomposition factors to avoid effects from cache locality.
Figure~\ref{fig:weak-scaling:cpu} shows the weak scaling results. We see that overdecomposition reduces the runtime by more than 50\% in some cases.

For the experiment on GPUs, we used the g4dn.xlarge instance type with 4 vCPUs each and 1 NVIDIA\textsuperscript{\textregistered} T4 Tensor Core GPU without RDMA support, and a grid size of $16,384 \times 16,384$ per GPU. Figure~\ref{fig:weak-scaling:gpu} shows the result of weak scaling on GPUs. We see that on a single GPU, overdecomposition increases the average time per iteration since there is no latency hiding to offset the overhead of overdecomposition. As we increase the number of GPUs, an overdecomposition factor of 4 performs consistently better than no overdecomposition and shows good weak scaling performance even without GPUDirect RDMA.

These results show that even at higher overdecomposition factors, the performance loss is minimal for both CPU and GPU applications. Thus, a larger overdecomposition factor is a worthwhile tradeoff to enable load balancing for applications that are dynamic or running on heterogeneous hardware.

\subsection{Processor Heterogeneity and Variability}
\label{sec:hetero}

Demand for cloud instances can vary significantly, and during peak times, the capacity in a given instance pool can be exhausted. To obtain the required capacity, users often need to source instances from different availability zones or other instance pools. However, instances in different availability zones are unsuitable for HPC applications due to high communication latency. Consequently, HPC applications on the cloud must often use instances from multiple pools, which typically consist of VMs running on heterogeneous hardware.

Moreover, the multi-tenant nature of cloud resources can cause performance variability due to CPU sharing. Thus, an adaptive runtime system that can efficiently balance loads across heterogeneous processors and accelerators is essential for running large-scale HPC on the cloud.
The rate-aware dynamic load balancing capability of Charm++ makes it an ideal choice for building HPC applications for execution on heterogeneous hardware.

\subsubsection{Experiments}

To study the effect of rate-aware dynamic load balancing of Charm++ on CPU applications running on heterogeneous processors, we use two benchmarks - Jacobi2D, a communication-intensive benchmark, and LULESH, a computation-intensive unstructured Lagrangian shock hydrodynamics proxy application~\cite{lulesh, lulesh-ref}. Both these applications have no load imbalance, and on homogeneous processors should run most efficiently without load balancing. We ran these applications on an EC2 fleet with a diversified spot allocation strategy. Table~\ref{tab:hetero-instance-details} shows the fleet configurations used in the experiments.

\begin{table}[h]
\centering
\caption{Diversified instance configurations for heterogeneity experiments.}
\label{tab:hetero-instance-details}
\begin{tabular}{@{}lll@{}}
\toprule
\textbf{Category} & \textbf{Instance Types} & \textbf{Hardware Details (per Instance)} \\ \midrule
\textbf{CPU Fleet} & \makecell[l]{c7i, c6a, c5a, c5d \\ (xlarge / 2xlarge)} & Intel/AMD x86 (4--8 vCPUs) \\ \midrule
\textbf{GPU Fleet} & g5.2xlarge & 1x NVIDIA A10G (24 GB) \\
                   & g6e.2xlarge & 1x NVIDIA L40S (48 GB) \\ \bottomrule
\end{tabular}
\end{table}

\begin{figure*}
    \centering
    \begin{subfigure}[b]{.32\linewidth}
        \centering
        \includegraphics[width=\linewidth]{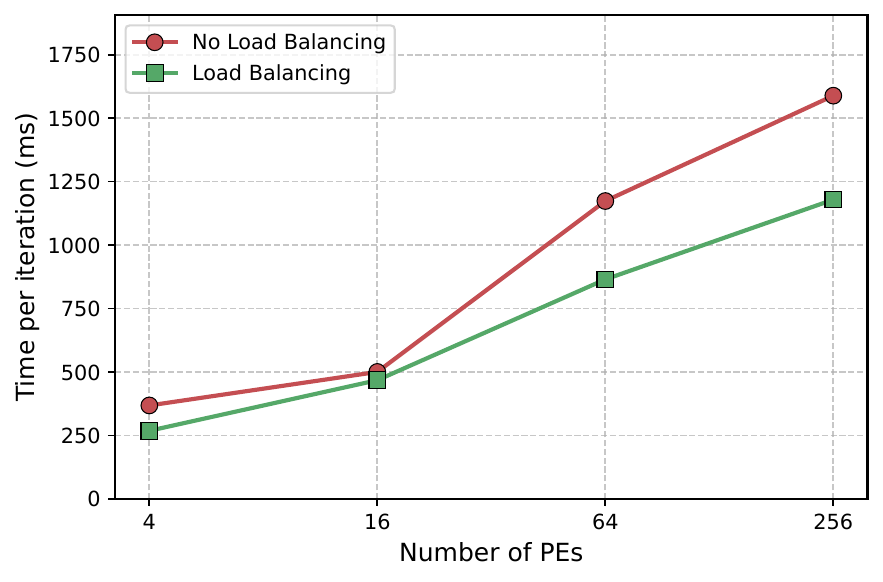}
        \caption{Weak scaling for Jacobi2D on CPUs}
        \label{fig:hetero:jacobi}
    \end{subfigure}
    \hfill
    \begin{subfigure}[b]{.32\linewidth}
        \centering
        \includegraphics[width=\linewidth]{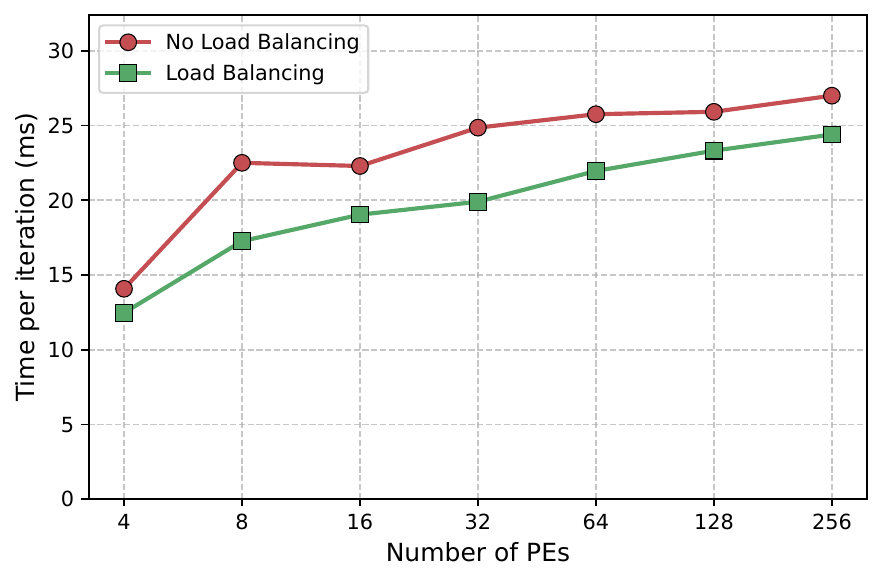}
        \caption{Weak scaling for LULESH on CPUs}
        \label{fig:hetero:lulesh}
    \end{subfigure}
    \hfill
    \begin{subfigure}[b]{.32\linewidth}
        \centering
        \includegraphics[width=\linewidth]{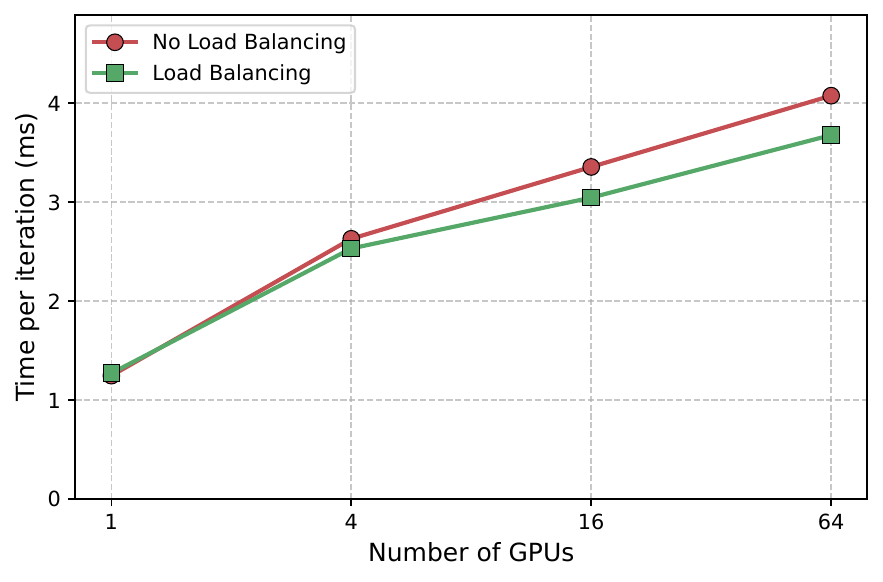}
        \caption{Weak scaling for Jacobi2D on GPUs}
        \label{fig:hetero:jacobi-gpu}
    \end{subfigure}
    \caption{Weak scaling with and without load balancing on heterogeneous CPUs and GPUs}
    \label{fig:hetero}
\end{figure*}

For CPU experiments we used the c7i, c6a, c5a, and c5d instance types. We chose x86 instances for these experiments since they offer greater hardware diversity than the ARM\textsuperscript{\textregistered} instances. We used the xlarge instance size for up to 64 PEs, switching to the 2xlarge size for larger PE counts. We used the GreedyRefine load balancing strategy from the Charm++ suite of load balancers for both benchmarks. This strategy attempts to keep most objects to their current home, aiming to minimize object migration and disruption of communication locality.
Figures~\ref{fig:hetero:jacobi} and~\ref{fig:hetero:lulesh} show the weak scaling results for these benchmarks.

For Jacobi2D, rate-aware load balancing improved performance by approximately 26\% on 256 PEs compared to no load balancing.
In the case of LULESH, we see that rate-aware load balancing consistently improved performance by 10\% to 25\%.

To study the effect of rate-aware dynamic load balancing on GPU applications, we ran the Jacobi2D application on an EC2 fleet of g5.2xlarge and g6e.2xlarge instances with NVIDIA\textsuperscript{\textregistered} A10G and NVIDIA\textsuperscript{\textregistered} L40S GPUs respectively, with a diversified spot allocation strategy. We ran a weak scaling experiment with a grid size of $16,384 \times 16,384$ per GPU. Figure~\ref{fig:hetero:jacobi-gpu} shows the average time per iteration with and without load balancing. We see an improvement of approximately 10\% in performance from rate-aware load balancing.

\section{HPC on Spot Instances}

Cloud providers offer spare, discounted capacity through spot instances, which can be preempted with a short warning when demand for on-demand instances surges. Previous work utilizing the \texttt{CharmCloudManager} framework to run HPC applications on spot instances with Charm++ demonstrated significant cost savings~\cite{Bhosale2025Cloud}. However, that work was limited to CPUs and revealed significant overhead from rescaling in response to an interruption. Furthermore, the transient loss of capacity from a spot interruption substantially impacted the overall runtime. In this paper, we extend \texttt{CharmCloudManager} to support GPUs and introduce a capacity rebalancing feature. This feature proactively launches replacement instances when an instance is at risk of interruption, thereby minimizing the transient drop in capacity. We also compare the performance of Charm++'s in-memory checkpointing based handling of spot interruptions with a traditional filesystem-based approach.

\subsection{GPU Support}
\label{sec:spot:gpu}

While in-memory checkpointing is an efficient approach to handling spot interruptions on CPUs, using the same approach for handling interruptions in GPU applications would require GPU-resident data to be copied from device to host at checkpoint and then back from host to device on restart. To avoid the data movement between the host and device during interruptions, we checkpoint GPU-resident data on the device using a daemon process~\cite{Bhosale-sci25}.

We launch a daemon process for each GPU on each instance in our fleet at startup. When an instance is to be interrupted, the runtime system first migrates the chares residing on that instance to other instances. For checkpointing the application state, the CPU data on each instance is checkpointed in Linux\textsuperscript{\textregistered} shared memory as usual, and the GPU-resident data pointers are sent to the daemon process of the corresponding device using CUDA Interprocess Communication (IPC). The daemon copies the data to a new buffer allocated by the daemon process. After restart, the daemon processes send the pointers to the GPU data back to the Charm++ PEs, and the data is copied back into the chare owned device buffers. 

When a replacement instance is launched, daemon processes are first launched on the replacement instance. The application is then checkpointed and restarted using the same protocol. After restart, the application is load balanced to distribute chares to the new replacement instances.

If a replacement for the instance to be interrupted is available at the time of the rescaling call, the chares from the interrupted instance are migrated to other instances. The application is then checkpointed and restarted with the new replacement instance. A subsequent load balancing step balances the load across all instances, avoiding an additional checkpoint-restart step.

\subsection{Capacity Rebalancing}
\label{sec:spot:rebalance}

\begin{figure*}
    \centering
    \includegraphics[width=\linewidth]{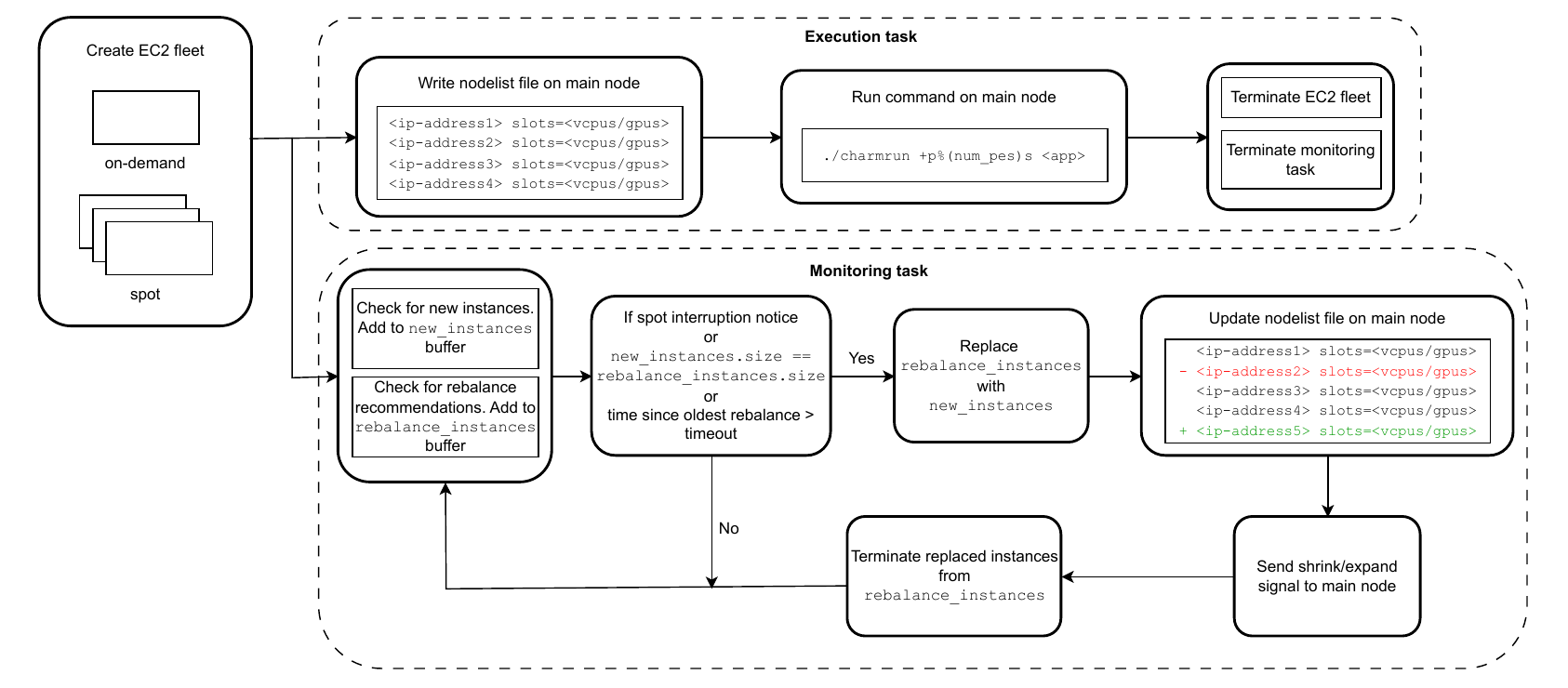}
    \caption{\texttt{CharmCloudManager} framework with capacity rebalancing}
    \label{fig:spot-rebalancing}
\end{figure*}

Previous work on running Charm++ on spot instances showed that the temporary drop in capacity can add a major overhead to the overall runtime of the application~\cite{Bhosale2025Cloud}. In this paper, we use the capacity rebalancing feature of EC2 fleets to address that issue. With capacity rebalancing, the instances at an elevated risk for interruption are sent a rebalance notification. On receiving a rebalance notification, the EC2 fleet proactively looks for and launches a replacement instance from the instance pools specified by the user. Figure~\ref{fig:spot-rebalancing} shows the new architecture for \texttt{CharmCloudManager} we developed to support this proactive replacement strategy.

The monitoring task continuously checks the fleet, identifying newly launched instances and tracking instances that receive a rebalance recommendation. To manage the replacement process efficiently, the system waits to signal the application to rescale its workload until one of three conditions is met.

The first and ideal condition is a complete replacement, where the rescale is triggered after all instances with a rebalance recommendation have been successfully replaced.

The second condition is an emergency override: a rescale is triggered immediately if any at-risk instance receives a spot interruption notice. This initiates a partial replacement, where any available new instances are prioritized to replace the instance with the interruption notice.

The third condition is a timeout: a rescale is triggered once $T_{timeout}$ seconds have elapsed since the oldest rebalance recommendation was received. This policy ensures that ready replacement instances sit idle for at most $T_{timeout}$ seconds while waiting for the entire replacement cycle to complete. 

In both the emergency and timeout scenarios, any at-risk instances that were not replaced continue to run in the fleet, and the application rescales using the capacity that is currently available.

The choice of $T_{timeout}$ involves a trade-off between overhead and cost. For deadline-sensitive applications, a larger $T_{timeout}$ is ideal because it minimizes instance replacement overhead by delaying rescaling, which increases the likelihood of replacing multiple instances in a single operation. Conversely, to minimize application costs while avoiding a drop in capacity, a smaller $T_{timeout}$ should be used. This reduces the cost of idle replacement instances by triggering the rescale earlier. In our experiments, we used a timeout of 120 seconds.

\subsection{Experiments}

In this section, we conduct three experiments to study the effect of spot interruptions on application performance. First, we evaluate the shrink and expand overheads when a CPU application encounters an interruption. Second, we measure this same overhead for a GPU application. Finally, we compare the interruption overheads and application performance across three modes of handling interruptions.
Table~\ref{tab:fault-injection-details} shows the instances used for the experiments in this section.

\begin{table}[h]
\centering
\caption{Instance configurations for spot interruption and rescaling overhead experiments.}
\label{tab:fault-injection-details}
\begin{tabular}{@{}lll@{}}
\toprule
\textbf{Category} & \textbf{Instance Type} & \textbf{Hardware Details (per Instance)} \\ \midrule
\textbf{CPU} & c6gn.large & 2 vCPUs (AWS Graviton2) \\ \midrule
\textbf{GPU} & g4dn.xlarge & 4 vCPUs, 1 NVIDIA T4 GPU (16 GB) \\ \bottomrule
\end{tabular}
\end{table}

\subsubsection{Interruption overhead on CPUs}

\begin{figure*}
    \centering
    \includegraphics[width=.8\linewidth]{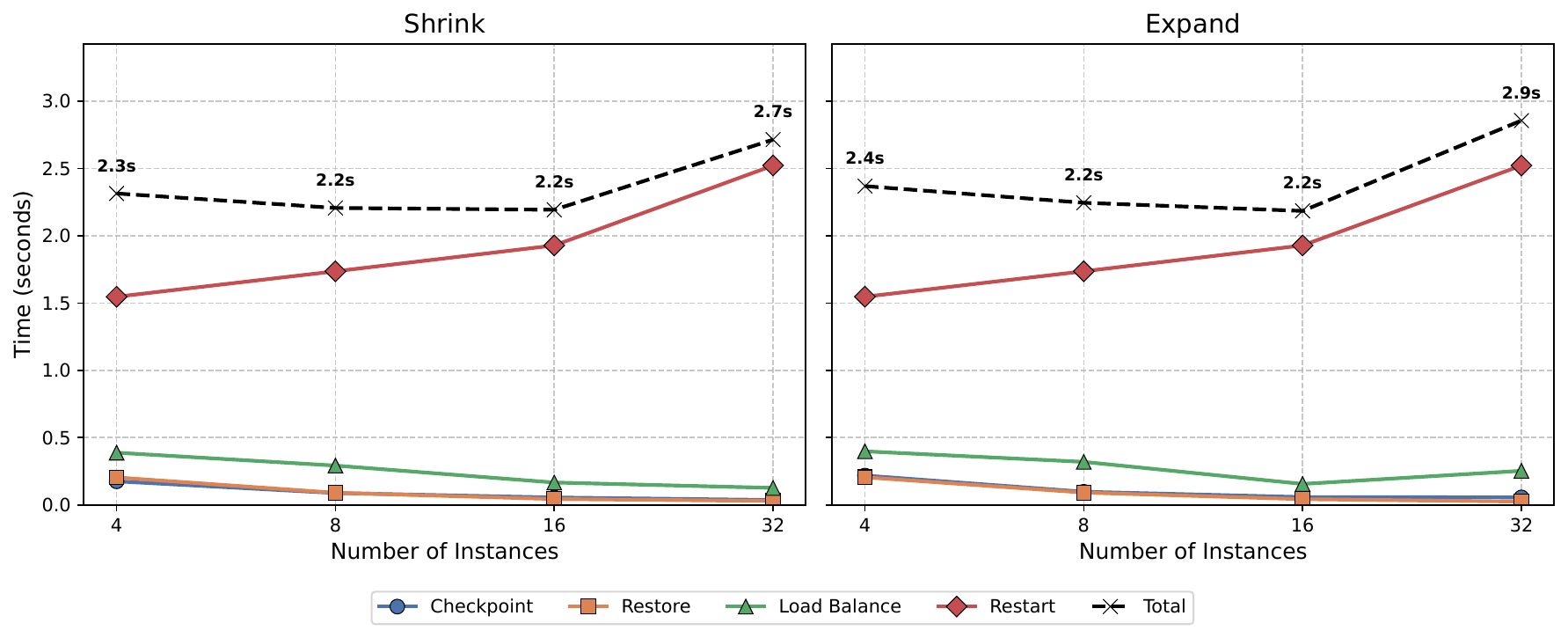}
    \caption{Overhead of spot interruption on CPUs when one instance is interrupted}
    \label{fig:overhead-cpu}
\end{figure*}

For this experiment, we used an EC2 fleet of c6gn.large instances, each with 2 vCPUs. We ran the Jacobi2D application with a grid size of $16,384 \times 16,384$ and injected a spot interruption into the fleet using the AWS Fault-Injection Simulator (FIS). We measured two primary overheads: the time to scale down the application upon receiving an interruption notice, and the time to scale up once a replacement instance was launched. Figure~\ref{fig:overhead-cpu} shows these overheads broken down into the four stages of rescaling: checkpoint, load balance, restart, and restore~\cite{Bhosale2025Cloud}.

We see that the time taken to checkpoint, restore, and load balance reduces as the number of instances is increased since the amount of data per instance reduces. 
The restart time increases with the number of instances since the application startup time increases with the number of nodes.
The total overhead is dominated by the restart time for both shrink and expand.

\subsubsection{Interruption overhead on GPUs}

\begin{figure*}
    \centering
    \includegraphics[width=.8\linewidth]{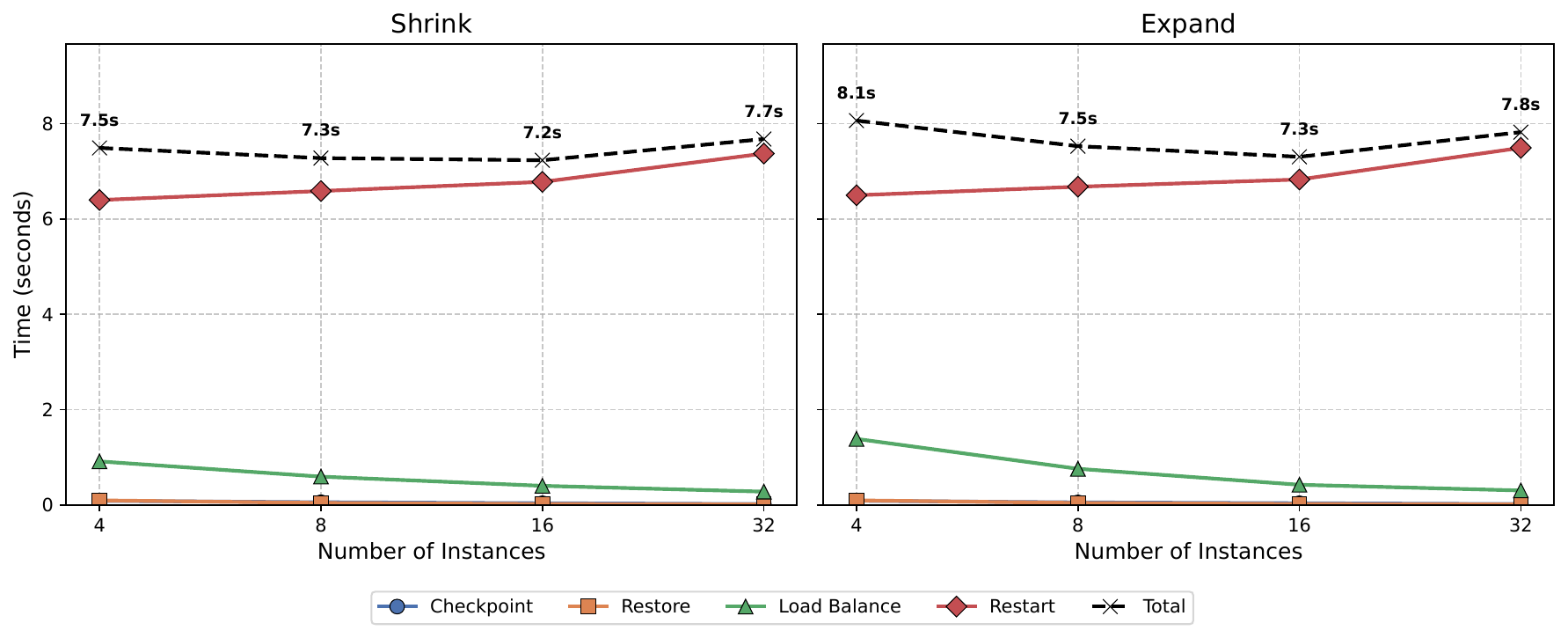}
    \caption{Overhead of spot interruption on GPUs when one instance is interrupted}
    \label{fig:overhead-gpu}
\end{figure*}

For this experiment, we used an EC2 fleet of g4dn.xlarge instances. We ran the GPU implementation of the Jacobi2D application with a $16,384 \times 16,384$ grid. Similar to the CPU experiment, we used AWS FIS to inject a spot interruption and measured the resulting shrink and expand overheads. These overheads, broken down into four stages, are shown in Figure~\ref{fig:overhead-gpu}.

We see that the checkpoint and restore times are significantly lower than the CPU for the same problem size. This is because the checkpointing of GPU data is done on the daemon process, which essentially copies the data into a different buffer on the same GPU. Since the bandwidth of the GDDR6 memory used on the T4 GPU is significantly higher than the bandwidth of the DDR4 memory on the AWS Graviton2 processor used in the c6gn.large instance, the overhead of checkpoint and restore is much lower for the former.

We see that the load-balancing time is higher for the GPUs than for the CPUs. This is because the GPU-resident data has to migrate to other devices via host staging since this instance type does not support GPUDirect RDMA, which adds extra overhead for migration. However, if an RDMA enabled instance is used, Charm++ will automatically use GPUDirect for migrating GPU-resident data.

The restart overhead of both shrink and expand on GPUs is significantly greater than on CPUs due to the additional CUDA initialization overhead. We found that the launch of daemon processes on each instance did not add a significant overhead to the startup time for the scale of our experiments.

\subsubsection{Comparison of interruption handling modes}

In this experiment, we evaluate and compare three different modes of handling spot interruptions. The first is the commonly used method for MPI applications: checkpointing to disk and restarting after an interruption~\cite{Yi10, Yi12}. Checkpointing to disk, however, is an expensive operation and requires a shared filesystem attached to each instance, which incurs an additional monetary cost. The second mode we evaluate is the mechanism described in Bhosale et al.~\cite{Bhosale2025Cloud}, which uses the spot interruption notice without proactive instance replacement. Finally, the third mode is our proposed mechanism of using capacity rebalancing to launch replacement instances proactively.

\begin{figure*}
    \centering
    \begin{subfigure}[b]{.45\linewidth}
        \centering
        \includegraphics[width=\linewidth]{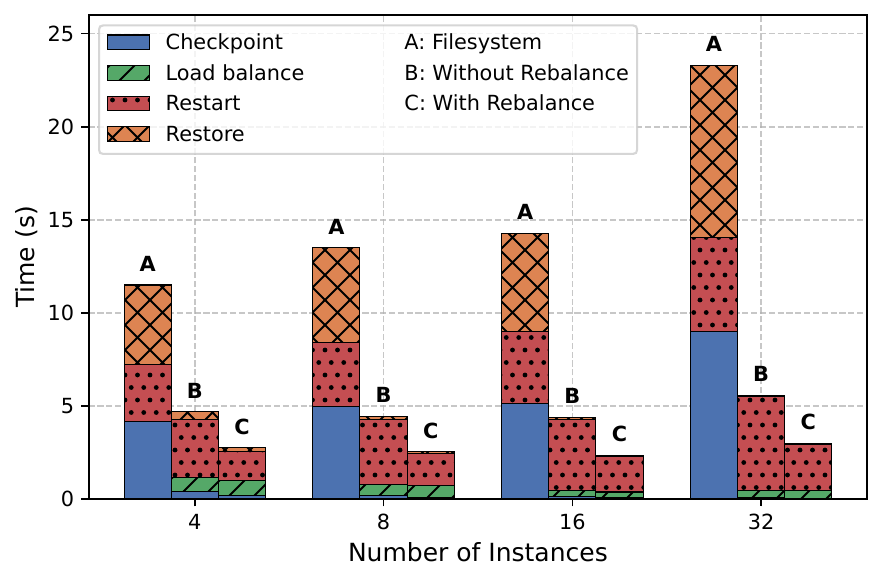}
        \label{fig:interruption-modes:cpu}
        \caption{Overhead comparison on CPUs}
    \end{subfigure}
    \begin{subfigure}[b]{.45\linewidth}
        \centering
        \includegraphics[width=\linewidth]{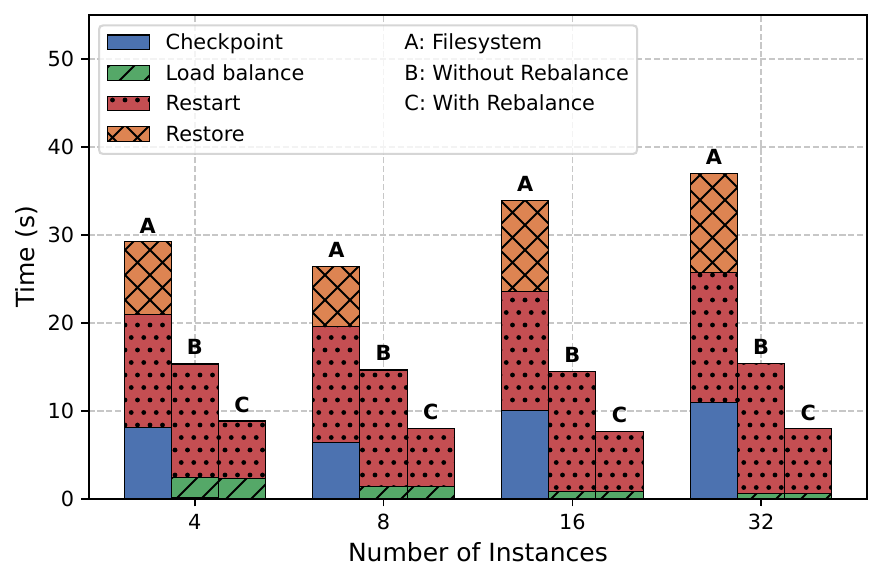}
        \label{fig:interruption-modes:gpu}
        \caption{Overhead comparison on GPUs}
    \end{subfigure}
    \caption{Comparison of overhead with 3 modes of handling spot interruptions when one instance is interrupted. \textit{Mode A}: Using a shared filesystem to checkpoint the application state. \textit{Mode B}: Using in-memory checkpointing without capacity rebalancing. \textit{Mode C}: Using in-memory checkpointing with capacity rebalancing.}
    \label{fig:interruption-modes}
\end{figure*}

To compare the 3 modes of handling spot interruptions, we first repeat the previous two experiments using the 3 handling modes.
For the shared filesystem checkpointing mode, we used AWS Elastic File System (EFS) with elastic throughput.
Figure~\ref{fig:interruption-modes} shows the total overhead of a spot interruption for CPUs and GPUs using each of the 3 modes.

The shared filesystem checkpointing mode consists of three stages: checkpoint, restart, and restore. The overhead for both checkpointing and restoring to a shared filesystem is significantly larger than for in-memory checkpointing. Furthermore, unlike the in-memory approach, the overhead of filesystem-based checkpointing increases as the number of instances grows.

We see that in-memory checkpointing incurs less than 50\% of the overhead of using filesystem-based checkpointing in most cases on both, CPUs and GPUs.
Using capacity rebalancing further reduces the total interruption handling overhead by approximately 50\%. This reduction is primarily attributed to a lower restart time, as capacity rebalancing allows the application to restart only once instead of twice following an interruption. Although the checkpoint and restore times are also halved, their contribution to the total overhead is less significant due to their comparatively small magnitude.

\begin{figure*}[h!]
    \centering
    \begin{subfigure}[b]{.45\linewidth}
        \centering
        \includegraphics[width=\linewidth]{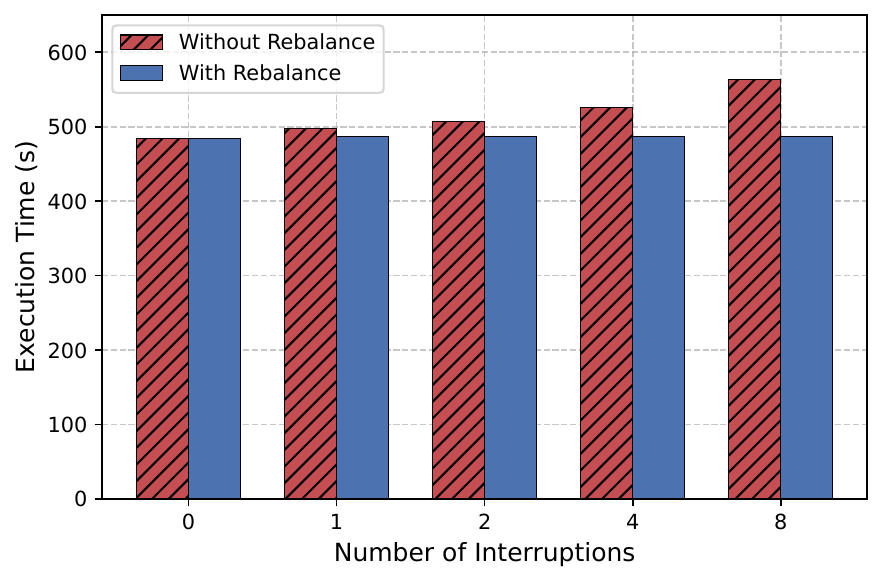}
        \label{fig:rebalance-e2e:cpu}
        \caption{Total time on CPUs}
    \end{subfigure}
    \begin{subfigure}[b]{.45\linewidth}
        \centering
        \includegraphics[width=\linewidth]{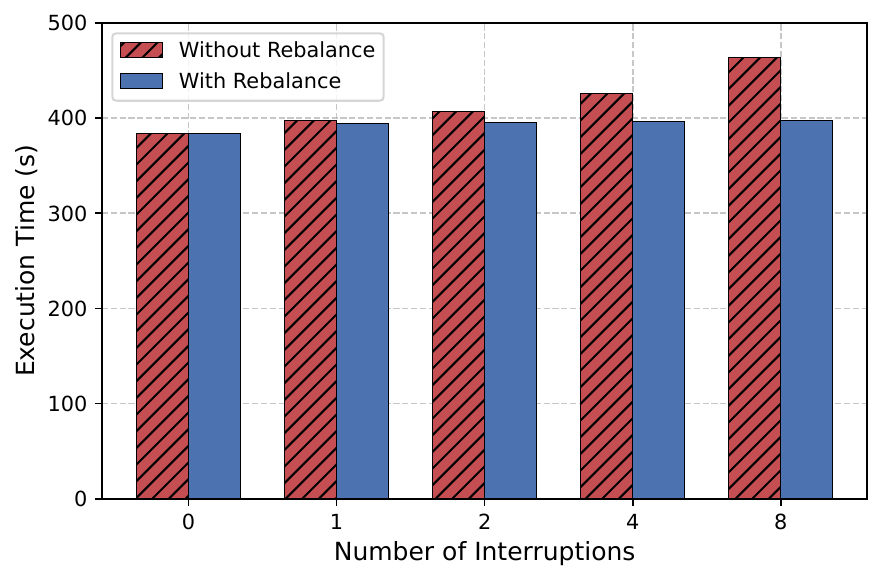}
        \label{fig:rebalance-e2e:gpu}
        \caption{Total time on GPUs}
    \end{subfigure}
    \caption{Total time for Jacobi2D application running on 16 instances with different numbers of interruptions}
    \label{fig:rebalance-e2e}
\end{figure*}

To study the effect of capacity rebalancing on the total runtime of an application, we measure the end-to-end runtime of Jacobi2D on 16 instances, with a different number of instances being interrupted simultaneously. For CPUs, we used the c6gn.large instance type with a grid size of $16,384 \times 16,384$ and $5000$ iterations. For GPUs, we used the g4dn.xlarge instance type with a grid size of $32,768 \times 32,768$ and ran for $30,000$ iterations.
Figure~\ref{fig:rebalance-e2e} shows the total runtime for the application on both CPUs and GPUs. 

We see that the total runtime without rebalancing increases by approximately 16\% on CPUs and by 20\% on GPUs when 8 instances are interrupted because the application has to run at a lower capacity until the replacement instances are launched, and because of the overhead of two rescaling operations. However, with capacity rebalancing, the runtime increases by less than 1\% on CPUs and by 3.4\% on GPUs when 8 instances are interrupted since the application runs at its target capacity for the entire duration of the run, and the overhead of rescaling is further reduced by using a single rescaling call to handle interruptions.

\section{Related Work}

Several papers have studied the performance of MPI on the cloud and proposed optimizations to improve communication performance~\cite{Charkraborty19, Xu20, Xu22}.
Sochat et al.~\cite{sochat-arxiv25} conducted a cross-platform usability study assessing common HPC proxy applications across AWS, Azure, GCP, and on-premise HPC clusters, and presented performance data on up to 256 GPUs.

Existing research on running HPC applications on spot instances typically uses checkpoint-restart mechanisms to handle interruptions~\cite{Yi10, Yi12}. These studies primarily rely on disk-based checkpointing and focus on adaptive strategies that adjust checkpointing frequency based on spot price fluctuations. By using time-series analysis of historical price data, these methods dynamically decide when to trigger a checkpoint to minimize the loss of progress while keeping the total checkpoint overhead low.

Another approach for handling spot interruptions in MPI applications combines checkpoint-restart with replicated execution~\cite{Marathe15, Gong15}. This strategy was particularly relevant when spot prices were determined by bidding. Replicated execution leverages price disparities across availability zones and instance types. It allows an application to checkpoint in a zone where the spot price is approaching the bid threshold and resume computation on an equivalent allocation in a more cost-effective zone.

Unlike these MPI-based approaches, which often have to deal with rigid, non-migratable ranks and external filesystem dependencies, the \texttt{CharmCloudManager} framework, first introduced in~\cite{Bhosale2025Cloud} and expanded in this work, leverages the migratable object model of Charm++ to handle volatility natively. While traditional methods are burdened by the I/O overhead of disk-based checkpointing, our approach utilizes in-memory and daemon-based checkpointing to minimize recovery time. Furthermore, we provide a mechanism that natively integrates with cloud-native signals like rebalance notifications, allowing the application to adapt dynamically rather than just reacting to failure.

While earlier work~\cite{Yi10, Yi12, Marathe15, Gong15} focused on the bidding-based auction models, more recent studies address the modern capacity-driven spot market. For instance, Wu et al.~\cite{wu-nsdi24} propose a Uniform Progress policy that moves away from price-prediction models. Instead, it uses a deadline-aware approach that dynamically switches to on-demand instances only when spot availability is insufficient to maintain a linear progress rate toward the deadline. This scheduling policy could complement \texttt{CharmCloudManager} for deadline-sensitive applications. Rather than relying on a static mix of spot and on-demand instances, this integration would further reduce deadline breaches caused by capacity drops as a result of low spot instance availability with minimal cost overhead.

Another avenue of research focuses on cloud-native resource management and scheduling. To reduce the overhead of running MPI on Kubernetes, several works have implemented custom schedulers to optimize pod placement and resource allocation~\cite{Misale2021, kube-batch, MPIOperator, flux}. Furthermore, newer elastic schedulers capitalize on cloud flexibility to dynamically scale HPC and AI/ML workloads at runtime~\cite{Bhosale-canopie25, voda}.

\section{Conclusion}

The convergence of HPC and cloud computing has created a need for runtime systems that can bridge the gap between applications designed for static, homogeneous supercomputers and the dynamic, heterogeneous nature of the cloud. In this paper, we demonstrated that the asynchronous, message-driven paradigm of Charm++ provides a powerful solution to this challenge.

We made two key contributions in this paper to create a more robust framework for cloud-native HPC. First, we demonstrated that core Charm++ features, such as support for overdecomposition, can be leveraged to successfully hide network latency, enabling efficient scaling on standard cloud interconnects. Furthermore, our results show that rate-aware dynamic load balancing effectively mitigates performance variability and enables efficient execution on heterogeneous CPUs and GPUs.

Second, we addressed the challenge of resilience and cost-effectiveness by improving the \texttt{CharmCloudManager} framework for volatile spot instances. We extended its capabilities to manage mixed fleets of on-demand and spot GPU instances and introduced a novel rebalancing strategy. This proactive approach to handling spot interruptions was shown to reduce the total overhead by approximately 50\% compared to the previous reactive approach of launching replacements after the interrupted instances were killed. We also showed that in-memory checkpointing for handling spot interruptions is vastly superior to the traditional filesystem-based approach. With our rebalancing strategy, the end-to-end application runtime increased by less than 1\% on CPUs and by 3.4\% on GPUs, even when half the instances in the run were interrupted.

\section{Acknowledgements}

This work is supported by the IBM-Illinois Discovery
Accelerator Institute (IIDAI).

\bibliographystyle{plain}
\bibliography{citations, group}

\end{document}